\documentclass[pra,twocolumn,preprintnumbers,amsmath,amssymb,nofootinbib,floatfix]{revtex4}
\usepackage{graphicx,bm}
\usepackage{float}
\usepackage{subfigure}
\usepackage{amsmath}
\usepackage{amsfonts}
\usepackage{tikz}
\usetikzlibrary{shapes.geometric, calc}
\usepackage{amsthm, amssymb,bbm,bm}
\makeatletter

\def\graphicscale{\twocolumn@sw{0.3}{0.4}}
\def\graphicthreescale{\twocolumn@sw{0.3}{0.4}}

\newcommand{\mc}[1]{\mathcal{#1}} 

\newcommand{\dt}[1]{\frac{d #1}{d t}}

\newcommand{\be}{\begin{equation}}
\newcommand{\ee}{\end{equation}}

\begin{document}
\title{Invariant Neural Network Ansatz for weakly symmetric Open Quantum Lattices}
\author{Davide Nigro}
\affiliation{Dipartimento di Fisica, Universit\`{a} di Pavia - via Bassi 6, I-27100 Pavia, Italy}
\begin{abstract}
We consider an open quantum system of $N$ interacting spin-1/2 particles on a $d$-dimensional lattice whose time evolution is (weakly) symmetric under the action of a finite group $G$ that is a subgroup of all the possible permutations of the lattice sites. We define a neural network representation for the system density operator that explicitly accounts for the system symmetries, and we show that its properties are preserved when using variational techniques such those recently used in the optimization of Restricted-Boltzmann Machines (RBMs). As a proof of principle, we demonstrate the validity of our approach by determining the steady state structure of the one dimensional dissipative XYZ model in the presence of a uniform magnetic field and uniform dissipation. We first show a direct comparison between the results obtained by optimizing a RBM representation for the density operator and those obtained by means of the invariant neural network defined in the present paper, and then we show results concerning the convergence to the unique steady state configuration when using our approach only. Numerical results support the conjecture that the invariant Ansatz defined in this paper is more expressive that a standard RBM and can be optimized with lesser numerical efforts. 
\end{abstract}
\maketitle
\section{Introduction}
In realistic conditions, quantum systems suffer a time evolution that is prominently conditioned by the coupling to the surrounding environment, and, in the most general case, an efficient description of such dynamics relies on the precise knowledge at every instant of the entire past of the system-environment correlations \cite{devega,nonmarkov2}. The description gets considerably simplified in the Markovian limit. Indeed, in the case of a weak interaction with a Markovian environment, the time evolution of an open quantum system is determined by the Lindblad--Gorini--Kossakowski--Sudarshan (LGKS) master equation \cite{Lindblad,Gorini,petruccionebook}, whose expression reads ($\hbar=1$):
\begin{equation}\label{eq:LGKS_equation}
\dt {\hat{\rho}} = \mc L [\hat{\rho}]= -i \left[\hat{\mc H},\,\hat{\rho}\right]+\mathbb{D}[\hat{\rho}]\\
\end{equation}
with $\hat{\rho}\equiv\hat{\rho}(t)$ representing the system density operator at time $t$. The global non-unitary dynamics encoded into Eq. \ref{eq:LGKS_equation} is determined by the competion of the unitary and purely dissipative processes captured respectively by the commutator with the Hamiltonian operator $\hat{\mc H}$ and by the \emph{dissipator} superoperator 
\be\label{eq:dissipator}
\mathbb{D}[\hat{\rho}]=\sum_{j}\gamma _j \left[L_j\hat{\rho} L^{\dagger}_j-\frac{1}{2}\left\{L_j^{\dagger}L_j,\,\hat{\rho}\right\}\right]
\ee
with $\left\{A,\,B\right\}\equiv AB + BA$ being the anticommutator of the operators $A$ and $B$, and with $\{L_j\}$ and $\{\gamma_j\}$ being the Lindblad operators and the corresponding decay rates respectively associated to a set of decay channels labeled by the index $j$.\\

The primary and most evident consequence of the lack of unitarity is that the LGKS equation leads to an \emph{irreversible} dynamics. In particular, with the passing of time, the system density operator $\hat{\rho}$ evolves towards one of the possible $N_{ss}$ \emph{steady state} configurations $\{\hat{\rho}^{(k)}_{ss}\}_{k=1,\cdots,N_{ss}}$, that is 
\begin{equation}\label{eq:steady_state}
\hat{\rho}^{(k)}_{ss} \equiv \lim_{t\to \infty} \hat{\rho}(t)
\end{equation} 
In the last decade, the characterization of such configurations has attracted a great deal of attention \cite{vestraete,prosenznidarich1,prosenznidarich2,geraceNat,znidarich1,prosen2011,ising_long,auffeves,diss_q1,lee2013,
XXZMendoza,prosen2014,bosehubbard,diss_q2,diss_q3,reservoirengineering1,
jin2016,diss_q4,bartolo2016,Mascarenhas2016,casteels2017,biella2017,biondi2017,savona2017,rota2017}. However, the exact determination of such asymptotic states, even by means of numerical techniques, still represents a challenging task. Indeed, since the dimension of the Hilbert space $D$ usually grows exponentially with the system volume and since density operators depend on a number of parameters proportional to $D \times D$, handling the differential system in Eq. \ref{eq:LGKS_equation} becomes unfeasible for large systems.\\
Nevertheless, over the years, several approximate approaches have been developed, improved and successfully exploited to capture the fundamental features of steady states, see e.g. \cite{orus}. In this paper we move forward on this research line by paying attention to a class of such algorithms, namely variational schemes that optimize neural network representations of density operators like those introduced in \cite{open1,open2,open3,open4} (see also e.g. \cite{ham1,ham2,ham3,ham4,ham5,ham6} for applications of neural network to the physics of hamiltonian systems). In such seminal papers it has been shown that Restricted Boltzmann Machines (RBMs) can be optimized by means of different variational schemes in order to efficiently determine the steady states of some prototypical lattice models that recently attracted a great deal of attention, namely the dissipative counterparts of the XYZ and the Ising models in the presence of uniform dissipation with decay channels proportional to lowering spin operators. Such models, like most of the lattice models used in the characterization of dissipative quantum phase transitions, belong to a particular class of lattice models, that is open systems invariant under the action of a (weak) symmetry group of permutations $G$ and supporting a unique steady state. In particular, it has been shown in \cite{DNsymm} that for such class of open quantum systems the unique steady state configuration can be determined by exploring only a relevant subspace $\mathcal{I}_G$ of the space of density operators, whose dimension is considerably smaller than $D\times D$. Therefore, combining such theoretical findings with the techniques introduced in \cite{open1,open2,open3,open4} could be extremely useful, especially when dealing with large systems.  
Here we show how to accomplish such task by introducing a neural network representation that $(i)$ efficiently parametrises states belonging to $\mathcal{I}_G$ and $(ii)$ that evolves within such manifold under the action of the methods presented in \cite{open1,open2,open3,open4}.

The paper is organized as follows. In Sec \ref{sec:variational_opt_RBM} we briefly summarize the ideas at the basis of the optimization strategies discussed in \cite{open1,open2,open3,open4}. In Sec \ref{sec:theory}, we first provide some details concerning the concept of weak symmetry in the context of dissipative systems, and then we show how to take advantage of the constraint imposed by the symmetry group by introducing a new neural network Ansatz. In particular, in this section we explicitly show that the variational optimization of such new representiation can be performed, without altering the symmetry properties, by means of \emph{any} of the variational schemes summarized in Sec \ref{sec:variational_opt_RBM}. In Sec \ref{sec:numerical_results} we show numerical results supporting our theoretical discussion. We first describe the model investigated numerically and its symmetries (Sec \ref{sec:model}), and then provide some details about the optimization strategy used in the present analysis (Sec. \ref{sec:evolution_method}). After these two preliminary sections, we show results obtained by simulating the dynamics of the one dimensional XYZ model in the presence of a uniform magnetic field and uniform dissipation. In particular, in Sec. \ref{sec:performance_comparison} we provide a direct comparison between the performances of a Restricted Boltzmann Machine while increasing the number of variational parameters and those of a neural network having the structure presented in this work. Here we show that if on the one hand the RBM representation becomes more expressive and provides a more accurate approximation of the steady state while increasing the number of variational parameters, on the other hand the invariant Ansatz is capable of capturing the steady state properties efficiently when using the \emph{minimum number} of variational parameters that can be used in the RBM representation. In Sec. \ref{eq:effective_time_evolution} we show numerical results proving the convergence of the invariant Ansatz to the actual steady state for other different values of the number of spins and in corresponcende of different points in the parameters space. Finally, in Sec. \ref{sec:summary_and_conclusions} we summarize our findings and draw our conlclusions.\\
\section{Variational optimization of a Restricted Boltzmann Machine}\label{sec:variational_opt_RBM}
Let us consider an open quantum system of $N$ spin-1/2 particles, whose dynamics is described by a master equation having the structure provided in Eq. \ref{eq:LGKS_equation}. In this case, the state of the system is described at any instant $t$ by a $2^N \times 2^N$ density operator, that without any loss of generality has the following representation
\begin{equation}
\hat{\rho}=\sum_{\bm{\sigma},\,\bm{\eta}} c_{\bm{\sigma}\bm{\eta}} \vert \bm{\sigma}\rangle\langle\,\bm{\eta}\vert, \quad c_{\bm{\sigma}\bm{\eta}}\in\mathbb{C}
\end{equation} 
with $\vert \bm{\sigma}\rangle = \bigotimes_{k=1}^{N}\vert \sigma_k \rangle $ denoting an eigenstate of the total magnetization along the $z$ direction, that is $\hat{M}_z=\sum_{k=1}^N\,\hat{\sigma}^{k}_{z}$, with 
\begin{equation}
\hat{\sigma}^{k}_{z}=\bm{1}_1\otimes \bm{1}_2 \otimes \cdots\bm{1}_{k-1}\otimes \hat{\sigma}_{z}\otimes\bm{1}_{k+1}\otimes \cdots\otimes \bm{1}_N,
\end{equation}
with $\hat{\sigma}_{z}$ denoting the Z-Pauli matrix and with $\sigma_k=\pm 1$ being the orientation along the $z$-axis of the $k$-th spin of the lattice. In particular, each steady state corresponds to a particular choice of the set of coefficients $\{c_{\bm{\sigma}\bm{\eta}}\}$ leading to a configuration annihilated by the super operator $\mc L$, that is
\begin{equation}\label{eq:steady_state_definition}
\mc L[\hat{\rho}^{(k)}_{ss}]=0,\quad \hat{\rho}^{(k)}_{ss}=\sum_{\bm{\sigma},\,\bm{\eta}} c^{(k)}_{\bm{\sigma}\bm{\eta}} \vert \bm{\sigma}\rangle\langle\,\bm{\eta}\vert
\end{equation}
In Refs. \cite{open1,open2,open3,open4} it has been shown that one can efficiently approximate steady state configurations by performing a variational optimization of trial states having the following structure
\begin{equation}\label{eq:rbm}
\hat{\rho}_{\vec{\chi}}=\sum_{\bm{\sigma},\,\bm{\eta}} \rho_{\vec{\chi}} (\bm{\sigma},\,\bm{\eta}) \vert \bm{\sigma}\rangle\langle\,\bm{\eta}\vert,
\end{equation}
with $\rho_{\vec{\chi}} (\bm{\sigma},\,\bm{\eta})$ being a function depending on a set of variational parameters corresponding to the components of the vector $\vec{\chi}$. In particular, independently of the particular variational scheme, the optimization procedure always aims at determining the values of the $N_p$ components of $\vec{\chi}$ that lead to an accurate approximation of the target steady state within some numerical threshold $\varepsilon > 0$. Notice, however, that it is not guaranteed \emph{a priori} that a given number of variational parameters is sufficient to meet some given numerical accuracy. Nevertheless, one can easily overcome such issue by increasing the \emph{expressive power} of the Ansatz, that is by enlarging the set of density operators efficiently represented by the neural network representation. Such task is achieved by simply varying the length of the variational vector $\vec{\chi}$. Indeed, results shown in Refs \cite{open1,open2,open3,open4} suggest that given some numerical threshold $\varepsilon$, there always exist a critical number of variational parameters $N^*_p$ such that 
\begin{equation}\label{eq:expressive_power}
\vert\vert \hat{\rho}^{opt}_{\vec{\chi}}\,-\, \hat{\rho}_{ss}\vert \vert=\sqrt{\sum_{\bm{\sigma},\,\bm{\eta}}\vert \rho^{opt}_{\vec{\chi}} (\bm{\sigma},\,\bm{\eta})-c^{(k)}_{\bm{\sigma}\bm{\eta}} \vert^2} < \varepsilon,\,\mbox{if}\, N_p > N^*_p
\end{equation}
with $\hat{\rho}^{opt}_{\vec{\chi}}$ denoting the optimal solution produced by the optimization technique when using $N_p$ variational parameters, and with $\vert\rho^{opt}_{\vec{\chi}} (\bm{\sigma},\,\bm{\eta})-c^{(k)}_{\bm{\sigma}\bm{\eta}}\vert$ denoting the absolute value of the difference between the elements in position $(\bm{\sigma},\,\bm{\eta})$ of the two matrices appearing in the left-hand side of Eq. \ref{eq:expressive_power}. In practice, one can icrease the number of variational parameters $N_p$ by increasing the number of hidden spins entering in the definition of the neural network Ansatz. In order to clarify this point, without any loss of generality, let us consider the RBM defined in Ref \cite{open1}. Such neural network Ansatz has the following matrix elements:
\begin{equation}\label{eq:rbm_elements}
\begin{split}
\langle \bm{\sigma} \vert \hat{\rho}_{\vec{\chi}}\vert \bm{\eta}\rangle &\equiv \rho_{\vec{\chi}} (\bm{\sigma},\,\bm{\eta})= 8 \exp(\sum_{j}a_j \sigma_j )\exp(\sum_{j}a^*_j \eta_j)\\
&\times \prod_{l=1}^{L}\mbox{cosh}\left(c_l+ \sum_{i}W_{li}\sigma_{i} + \sum_{i}W^*_{li}\eta_{i}\right) \\
&\times \prod_{m=1}^{M}\mbox{cosh}\left(b_m+ \sum_{i}X_{mi}\sigma_{i}\right) \\
&\times \prod_{n=1}^{M}\mbox{cosh}\left(b^*_n+ \sum_{i}X^*_{ni}\eta_{i}\right), \\
\end{split}
\end{equation}
with $\vec{a}=(a_1,\,a_2\,\cdots,a_N)$, $\vec{b}=(b_1,\,b_2\,\cdots,b_M)$ and $\vec{c}=(c_1,\,c_2,\cdots,c_L)$ representing three vectors of length $N$, $M$ and $L$ respectively, and with $W$ and $X$ denoting a $L \times N$  and a $M \times N$ matrix respectively. In this case, the components of the variational vector correspond to those of the vectors and matrices defined here above, that is 
\begin{equation}
\vec{\chi}\equiv(\{a_j\},\,\{b_n\},\,\{c_l\},\,\{W_{li}\},\,\{X_{ni}\}),
\end{equation}
and the number of complex parameters entering into $\hat{\rho}_{\vec{\chi}}$ is given by $N_p\equiv N+M+L+N(L+M)$. In particular, in this context, changing the number of hidden spins means to vary the parameters $M$ and $L$, or equivalently to vary the densities of hidden nodes that correspond to the following two real parameters
\begin{equation}
\alpha\equiv \frac{M}{N},\,\quad \beta\equiv \frac{L}{N}.
\end{equation} 
In other words, one can improve the expressive power of a RBM Ansatz such the one defined in Eq \ref{eq:rbm_elements} by properly increasing the length of the two vectors $\vec{b}$ and $\vec{c}$ and by increasing the number of rows of the two matrices $W$ and $X$ (please see e.g. Refs \cite{open1,open2,open3,open4} for further details).\\

Let us now pay attention to the optimization schemes. As shown in the seminal papers cited before, the numerical optimization of a trial state such the one defined in Eq. \ref{eq:rbm} can be performed by means of several iterative methods. To this purpose, let us consider the RBM Ansatz in a vectorized form, that is 
\begin{equation}
\vert \rho_{\vec{\chi}}\rangle=\sum_{\bm{\sigma},\,\bm{\eta}}\rho_{\vec{\chi}} (\bm{\sigma},\,\bm{\eta})\vert \bm{\sigma },\,\bm{\eta}\rangle, 
\end{equation}
and suppose the variational vector at the step $n$ of the iteration scheme to be denoted by $\vec{\chi}_{(n)}$. Our goal now is to determine the next variational vector in the sequence, that is $\vec{\chi}_{(n+1)}$. Since the dynamics drives the system configuration towards steady states, a possibile way to update the variational vector is to design the variational scheme in order to mimick the actual time evolution. Such task is performed, for instance, by minimizing the distance between the two following vectors
\begin{equation}\label{eq:update_chi}
\vert \rho_{\vec{\chi}_{(n)}+d\vec{\chi}}\rangle\approx\vert\rho_{\vec{\chi}_{(n)}}\rangle+ \sum_{k=1}^{N_p}d{\chi_k}O_k \vert\rho_{\vec{\chi}_{(n)}}\rangle,
\end{equation} 
and 
\begin{equation}\label{eq:dt_chi}
\vert\rho_{\vec{\chi}_{(n)}}(t+dt)\rangle\approx\vert\rho_{\vec{\chi}_{(n)}}(t)\rangle+dt\,M_{\mathcal{L}}\vert\rho_{\vec{\chi}_{(n)}}\rangle,
\end{equation}
with $M_{\mc L}$ denoting the matrix describing the action of $\mc L$ on $\vert \rho_{\vec{\chi}}\rangle$, and with
\begin{equation}\label{eq:ok_vectors}
O_k=\sum_{\bm{\sigma},\,\bm{\eta}}\frac{\partial}{\partial \chi_k} ln(\rho_{\vec{\chi}_{n}}(\bm{\sigma},\,\bm{\eta}))\vert \bm{\sigma},\,\bm{\eta}\rangle\langle\bm{\sigma},\,\bm{\eta}\vert
\end{equation}  
The state in Eq.\ref{eq:update_chi} denotes a variation of the RBM under an infinitesimal deformation of the variational vector $\vec{\chi}$, while the expression in Eq. \ref{eq:dt_chi} corresponds to a first order expansion in $dt$ of the time evolution of the RBM encoded into the LGKS equation. In particular, as shown in Refs. \cite{open1,open2}, the distance between such two vectors is minimized by updating the variational vector as follows
\begin{equation}
\vec{\chi}_{(n+1)}=\vec{\chi}_{(n)}+dt \frac{d\vec{\chi}}{dt}.
\end{equation}
with $d\vec{\chi}/dt$ being the solution of a linear system having the following form 
\begin{equation}\label{eq:linear_system}
S\frac{d\vec{\chi}}{dt}=\vec{f},
\end{equation}
with $S$ being a $N_p \times N_p$ matrix and $\vec{f}$ being a vector having $N_p$ components. As discussed in \cite{open2} both the structure of $S$ and $\vec{f}$ do depend explictly on the norm used in the minimization scheme. Nevertheless, by proceeding this way, the authors have shown that one can efficiently determine, within the expressive power provided by the Ansatz, which is the best sequence $\left\{\vec{\chi}_{(n)}\right\}$ of variational vectors that approximates the actual time evolution, thus leading in the vicinity of the steady state cofiguration. 

Another possible route, as shown in \cite{open3,open4}, relies on the minimization, by means of iterative gradient based schemes, of non-negative cost functions $\mc C (\vec{\chi})$ such the following ones
\begin{equation}\label{eq:ciuti_norm}
\mc C (\vec{\chi})=\frac{\langle \rho_{\vec{\chi}} \vert M^{\dagger}_{\mc L} M_{\mc L}\vert \rho_{\vec{\chi}} \rangle}{\langle  \rho_{\vec{\chi}}\vert  \rho_{\vec{\chi}}\rangle},
\end{equation}
and 
\begin{equation}\label{eq:toshioka_norm}
\mc C (\vec{\chi})=\langle \rho_{\vec{\chi}}\vert M^{\dagger}_{\mc L} M_{\mc L}\vert \rho_{\vec{\chi}} \rangle.
\end{equation}
Since when using the vectorized formalism steady states do correspond to vectors annihilated by $ M_{\mc L}$, that is
\begin{equation}
 M_{\mc L}\vert \rho^{(k)}_{ss}\rangle=0,
\end{equation}
and since the expressions in Eqs. \ref{eq:ciuti_norm} and \ref{eq:toshioka_norm} do describe non-negative functions, minimizing such quantities with respect to $\vec{\chi}$ is equivalent to determine which is the set of variational parameters that best approximate steady states. As a consequence, given a point $\vec{\chi}_{(n)}$ in the variational space and a cost function $\mc C (\vec{\chi})$ having steady states as minima, one can use the information encoded into the gradient vector of such latter quantity to determine the next variational vector in the sequence. For instance, when using the simplest gradient-based scheme, that is the steepest descent approach, the next variational vector in the sequence is chosen by moving in the opposite direction with respect to the gradient vector $ \left.\vec{\nabla}_{\vec{\chi}} \mc C(\vec{\chi})\right\vert_{\vec{\chi}_{(n)}}$, that is
\begin{equation}
\vec{\chi}_{(n+1)}=\vec{\chi}_{(n)}- d\nu \vec{\nabla}_{\vec{\chi}} \mc C
\end{equation}
with $0<d\nu\ll 1$ to ensure the convergence, and with 
\begin{equation}
\vec{\nabla}_{\vec{\chi}} \mc C=\left(\frac{\partial}{\partial\,\chi_1}\mc C,\,\frac{\partial}{\partial\,\chi_2}\mc C,\,\cdots,\,\frac{\partial}{\partial\,\chi_{\small{N_p}}}\mc C\right).
\end{equation}
Accordingly, the neural network representation is then updated in the following way
\begin{equation}\label{eq:update_chi_gradient}
\vert \rho_{\vec{\chi}_{(n)}} \rangle \rightarrow \vert \rho_{\vec{\chi}_{(n+1)}}\rangle \approx \vert\rho_{\vec{\chi}_{(n)}}\rangle- d\nu \sum_{k=1}^{N_p}(\vec{\nabla}_{\vec{\chi}} \mc C)_k O_k \vert\rho_{\vec{\chi}_{(n)}}\rangle.
\end{equation}
When using more sofisticated gradient-besad methods such those used in Refs \cite{open3,open4}, one determines the optimal way to update $\vec{\chi}_{(n)}$ in a fashion similar to what has been discussed in Ref \cite{open1,open2}, that is by using the solution $d\vec{\chi}/d\nu$ of a linear system like the one in Eq. \ref{eq:linear_system}, but having $\vec{f}=\vec{\nabla}_{\vec{\chi}} \mc C(\vec{\chi})$ to update the variational vector. Nevertheless, for our purposes it is worth noting that, independently of the particular optimization strategy used, the quantities exploited to update the structure of the neural network Ansatz depend only on linear combinations of $\vert \rho_{\vec{\chi}} \rangle$, $M_{\mc L}\vert \rho_{\vec{\chi}} \rangle$,  $M^{\dagger}_{\mc L}M_{\mc L}\vert \rho_{\vec{\chi}} \rangle$ and of the vectors $\{O_k \vert \rho_{\vec{\chi}} \rangle\}$. Indeed, one has that the $k$-th componentent of the gradients of the expressions in Eqs. \ref{eq:ciuti_norm} and \ref{eq:toshioka_norm} read 
\begin{equation}
\begin{split}
\frac{\partial}{\partial \chi_k}\langle\langle M^{\dagger}_{\mc L} M_{\mc L}\rangle\rangle&=\langle\langle O^{\dagger}_k M^{\dagger}_{\mc L} M_{\mc L}\rangle\rangle+\langle\langle M^{\dagger}_{\mc L} M_{\mc L} O_k\rangle\rangle\\
&-\langle\langle M^{\dagger}_{\mc L} M_{\mc L}\rangle\rangle\left[\langle\langle O^{\dagger}_k\rangle\rangle+\langle\langle O_k\rangle\rangle\right],\\
\end{split}
\end{equation}
and 
\begin{equation}
\frac{\partial}{\partial \chi_k}\langle M^{\dagger}_{\mc L} M_{\mc L}\rangle=\langle O^{\dagger}_k M^{\dagger}_{\mc L} M_{\mc L}\rangle+\langle M^{\dagger}_{\mc L} M_{\mc L} O_k\rangle
\end{equation}
respectively, with $\langle\langle \hat{A}\rangle\rangle\equiv \frac{\langle \rho_{\vec{\chi}} \vert\hat{A}\vert \rho_{\vec{\chi}} \rangle}{\langle  \rho_{\vec{\chi}}\vert  \rho_{\vec{\chi}}\rangle} $. Such fact plays a fundamental role in our discussion. Indeed, as shown in the following section, this implies that the variational optimization of our invariant Ansatz can be performed safely, that is without altering any of its symmetry properties, by means of any of the optimization strategies mentioned in this section.\\

\section{Theoretical framework and theoretical results}\label{sec:theory} 
In this section we define our invariant neural network and we show that its variational optimization can be performed by means of any of the variational schemes described in the previous section. We begin our discussion by observing that the models addressed in \cite{open1,open2,open3,open4} belong to a specific class of open quantum lattices, namely $d$-dimensional lattice systems whose dynamics is weakly invariant \cite{buca,albert} under the action of a subgroup $G$ of the permutation group, and having a unique steady state \cite{evans,daviesstochastic,schirmer,DNunique}. For such models, it has been shown in \cite{DNsymm} that the unique steady state can be determined \emph{exactly} by exploring only the subspace containing the density operators simultaneously commuting with all the elements of $G$, that is
\begin{equation}\label{eq:relevant_subspace}
\mathcal{I}_{G}=\left\{\hat{\rho} \,:\, \left[G_i,\,\hat{\rho}\right]= 0 \quad \forall\,G_i \in G \right\}
\end{equation}  
This fact follows directly from the weak invariance of the time evolution, that is 
\begin{equation}\label{eq:weak_invariance}
G_i\mc L[\hat{\rho}(t)]G_i^{\dagger}= \mc L[ G_i\hat{\rho}(t) G_i^{\dagger}],
\end{equation}
and from the uniqueness of the steady state, that is 
\begin{equation}
\lim_{t\to+\infty}\hat{\rho}(t)=\hat{\rho}_{ss}, \quad \forall \hat{\rho}_0\equiv \hat{\rho}(t=0)
\end{equation}
Indeed, whenever the steady state is unique, by combining the definition provided in Eq. \ref{eq:steady_state_definition} with the one in Eq. \ref{eq:weak_invariance}, one finds that 
\begin{equation}
\hat{\rho}_{ss}=G_i\hat{\rho}_{ss} G_i^{\dagger}\Leftrightarrow\,[G_i,\,\hat{\rho}_{ss}]=0
\end{equation}
which is equivalent to state that $\hat{\rho}_{ss}\in I_{G}$, see Eq. \ref{eq:relevant_subspace}.\\

As shown in Ref.\cite{DNsymm}, one can take advantage of the constraints imposed by the weak symmetry group by using as basis states the fully symmetrized configurations corresponding to projectors belonging to the same \emph{orbit}. More explicitly, instead of expanding the density operator in terms of the set of projectors $\{\vert \bm{\sigma}\rangle\langle\,\bm{\eta}\vert\}$, in the presence of a weak symmetry group such the one characterizing the dissipative models addressed in \cite{open1,open2,open3,open4} it is convenient to use as basis states a basis set of $I_G$, that is configurations having the following structure
\begin{equation}\label{eq:P_states}
P_{\bm{\sigma},\,\bm{\eta}}= \frac{1}{N_{\bm{\sigma},\,\bm{\eta}}}\sum_{G_i \in G}G_i\vert \bm{\sigma}\rangle\langle\,\bm{\eta}\vert G_i^{\dagger},
\end{equation}
with $N_{\bm{\sigma},\,\bm{\eta}}$ being a normalization factor.\\

In the present case, one can take advantage of such result by using the following \emph{invariant} neural network Ansatz 
\begin{equation}\label{eq:symmetrical_ansatz}
\hat{\rho}^{inv}_{\vec{\chi}}\equiv \frac{1}{\vert G \vert}\sum_{G_i \in G} G_i \hat{\rho}_{\vec{\chi}} G_i^{\dagger},
\end{equation}
with $\vert G \vert $ denoting the number of elements of $G$. By construction and, most importantly, independently of the particular form assumed for $\hat{\rho}_{\vec{\chi}}$, the Ansatz $\hat{\rho}^{inv}_{\vec{\chi}}$ commutes with all the elements of $G$. Therefore, it is well suited for targeting configurations belonging to $\mathcal{I}_G$ like the steady state. In addition, although the number of variational parameters used in the invariant representation $\hat{\rho}^{inv}_{\vec{\chi}}$ is \emph{exactly} the same number of variational parameters used in the $\hat{\rho}_{\vec{\chi}}$, numerical results shown in the following section suggest that the symmetrization procedure leads to a more expressive neural newtork representation. In order to understand this point, please consider the following set of projectors
\begin{equation}
\mc O (\bm{\sigma},\,\bm{\eta})=\left\{\,\vert \tilde{\bm{\sigma}}\rangle\langle\,\tilde{\bm{\eta}}\vert :\,\vert \tilde{\bm{\sigma}}\rangle\langle \tilde{\bm{\eta}}\vert=G_i\vert \bm{\sigma}\rangle\langle \bm{\eta}\vert G_i^{\dagger},\,G_i \in G\right\},
\end{equation}
which contains the projectors connected to $\vert \bm{\sigma}\rangle\langle\,\bm{\eta}\vert$ by a group transformation, that is its \emph{orbit}. Each density operator in $\mc I_G$, like the steady state, is such that    
\begin{equation}\label{eq:symmetry_prop}
\langle \bm{\sigma} \vert \hat{\rho}\vert \bm{\eta}\rangle =\langle \tilde{\bm{\sigma}} \vert \hat{\rho}\vert \tilde{\bm{\eta}}\rangle\quad \forall\, \vert \tilde{\bm{\sigma}}\rangle\langle\,\tilde{\bm{\eta}}\vert \in \mc O (\vert \bm{\sigma}\rangle\langle\,\bm{\eta}\vert)
\end{equation}
and the same property holds true by construction and independently of the number of variational parameters $N_p$ also for the invariant Ansatz, that is 
\begin{equation}\label{eq:matrix_elements_inv}
\rho^{inv}_{\vec{\chi}} (\bm{\sigma},\,\bm{\eta})=\rho^{inv}_{\vec{\chi}} (\tilde{\bm{\sigma}},\,\tilde{\bm{\eta}}).
\end{equation}
On the other hand this is not necessarily true for the neural network $\hat{\rho}_{\vec{\chi}}$ appearing into the right-hand side of Eq. \ref{eq:symmetrical_ansatz} in correspondence of an arbitrary number of variational parametes. As a consequence, in order to properly restoring the symmetries encoded into the steady state, when using a RBM Ansatz one would need to use more variational parameters. Therefore, one might expect that the symmetrization procedure leads, in general, to a more expressive neural network representation.\\

Furthermore, the variational optimization of the invariant Ansatz can be performed efficiently and without altering any of the properties listed here above by means of schemes such those briefly summarized before. This follows from the invariance of $\mathcal{I}_G$ under the action of $\mathcal{L}$ \cite{DNsymm}, and from the structure of the operators defined in Eq. \ref{eq:ok_vectors} for the invariant Ansatz. Indeed, due the property specified in Eq. \ref{eq:matrix_elements_inv}, one has that the vectors corresponding to an infinitesimal deformation of the invariant Ansatz, whose structure reads 
\begin{equation}\label{eq:ok_vectors_inv}
O^{inv}_k\vert \rho^{inv}_{\vec{\chi}}\rangle=\sum_{\bm{\sigma},\,\bm{\eta}}\frac{\partial}{\partial \chi_k} \rho^{inv}_{\vec{\chi}}(\bm{\sigma},\,\bm{\eta})\vert \bm{\sigma},\,\bm{\eta}\rangle,
\end{equation}
are invariant under the action of $G$. Therefore, any linear combination involving such vectors and our Ansatz, such the one considered in Eq. \ref{eq:update_chi} when using the schemes discussed in Refs. \cite{open1,open2} or the one considered in Eq. \ref{eq:update_chi_gradient} when exploiting methods similar to those used in Refs \cite{open3,open4}, always describes a state belonging to $\mathcal{I}_{G}$. In other words, we have that the symmetry properties encoded into the invariant Ansatz defined in Eq. \ref{eq:symmetrical_ansatz} are by construction preserved by any variational scheme similar to those discussed in Refs. \cite{open1,open2,open3,open4}.\\ 

\section{Numerical results}\label{sec:numerical_results}
In this section we corroborate our theoretical findings by showing some numerical results derived for the one dimensional dissipative XYZ model in the presence of a uniform magnetic field. As we discussed later, such lattice model belongs to the same class of systems addressed in Refs. \cite{open1,open2,open3,open4}. Nevertheless, before entering into detail, it is worth stressing that our analysis differs in several aspects from those presented in such papers. 
The primary goal of this paper is to understand if the symmetrization scheme described in Sec \ref{sec:theory} cuold lead to some improvements in the determination of the steady state configurations of open quantum lattices, with respect to those previously summarized. In particular, in this context, by improvement we mean any modification leading to a more precise determination of the steady state configuration at the same computational cost, that is an improvement in the expressive power of the neural network representation while keeping fixed the number of variational parameters, and/or leading to a reduction of the computational resources needed in the optimization scheme. For what concerns the former issue, numerical results supporting the conjecture that the invariant Ansatz introduced in this paper is more expressive than a standard RBM having the same number of variational parameters are shown in Sec. \ref{sec:performance_comparison}. For what concerns instead the latter issue, that is computational improvements, we cannot make a direct comparison between results obtained by means of our approach and those shown in Refs. \cite{open1,open2,open3,open4}, namely because we did not use stochastic methods. Indeed, results shown in the following sections have been derived by using full density operators, that is $2^N \times 2^N$ non-negative hermitian matrices, to compute both expectation values of observables as well as the many quantities involved in the variational optimization of the neural network representations. We opted for this approach because we were more interested in addressing the first issue. Nevertheless, on the basis of the discussion made in Refs. \cite{open1,open2,open3,open4} and on the numerical results shown in the following, we can make some general considerations suggesting that the symmetrization scheme introduced here can lead to computational improvements also when using Monte Carlo methods.\\

In general, at each iteration $n$ of the variational schemes discussed in \cite{open1,open2,open3,open4}, one first performs a sampling of the probability distribution associated to the neural network, and then has to find the solution of a linear system having size $N_p$ in order to update the variational vector $\vec{\chi}$. For what concerns the sampling strategy, for instance when considering a Metropolis update rule based on a probability $p(\bm{\sigma},\,\bm{\eta})\propto \vert\rho_{\vec{\chi}} (\bm{\sigma},\,\bm{\eta})\vert^2$, one can take advantage of symmetries by performing the following substitution 
\begin{equation}
\rho_{\vec{\chi}} (\bm{\sigma},\,\bm{\eta})\rightarrow \rho^{inv}_{\vec{\chi}} (\bm{\sigma},\,\bm{\eta}), 
\end{equation}
and use invariant configurations to compute expectations, that is
\begin{equation}
\vert \bm{\sigma}\rangle\langle\,\bm{\eta}\vert\rightarrow P_{\bm{\sigma},\,\bm{\eta}},
\end{equation}
with $P_{\bm{\sigma},\,\bm{\eta}}$ having the form in Eq. \ref{eq:P_states}. By doing so, each configuration $(\bm{\sigma},\,\bm{\eta})$ is properly weighted with all the others belonging to the same orbit, and the resulting Markov Chain is generated according to structure of an invariant density operator. Let us focus now on the second part of each iteration, that is finding the solution of a linear system of size $N_p$ to update $\vec{\chi}$. Suppose that using a RBM with $N_p$ parameters means to be capable of determining the steady state with a precision $\varepsilon$. In this context, having a more expressive Ansatz representation implies that the same level of accuracy $\varepsilon$ can be reached by using a smaller number of variational parameters, thus reducing the computational resources used at each iteration to update $\vec{\chi}$. Therefore, we do believe that using the symmetrization scheme discussed in this paper could lead to a significant reduction of the total computational cost needed for the variational determination of steady states also when using Monte Carlo methods.

\subsection{The model and its symmetries}\label{sec:model}
In our simulations we consider the time evolution of a system of $N$ spin-1/2 particles prescribed by the Lindblad master equation defined in Eq. \ref{eq:LGKS_equation}, with a Hamiltonian corresponding to the one dimensional XYZ in the presence of a uniform magnetic field and in the presence of a uniform dissipation along the chain. Furthermore, we assume periodic boundary conditions, that is $\sigma^k_{N+1}\equiv \sigma^k_1$, $k=x,\,y,\,z$. More explicitly, for what concerns the unitary part of the generator $\mc L$, in the present analysis we consider it as governed by the following Hamiltonian model ($\hbar = 1$)
\begin{equation}\label{eq:hamiltonian}
\hat{\mathcal{H}}=\sum_{i=1}^{N}\sum_{k=x,\,y,\,z}\left(J_{k} \sigma^k_i\sigma^k_{i+1}+B_{k} \sigma^k_i\right),
\end{equation}
with $J_{k}$ denoting the interaction strength between the $k$ component of spins placed on first-neighbour sites, and with $\vec{B}=(B_x,\,B_y\,B_z)$ denoting the local magnetic field. For what concerns the dissipative part of the Lindblad equation, we assume a uniform dissipation along the spin chain, that is
\be
\mathbb{D}[\hat{\rho}]=\sum_{j}\gamma _j \left[L_j\hat{\rho} L^{\dagger}_j-\frac{1}{2}\left\{L_j^{\dagger}L_j,\,\hat{\rho}\right\}\right]
\ee
with
\begin{equation}
\gamma_j\equiv \gamma,\quad L_j\equiv\sigma^{-}_j,\quad \sigma^-_j\equiv(\sigma^x_j+i \sigma^y_j)/2,\quad j\in[1,\,N]
\end{equation}
In the present case, due to the form of the dissipator, the Lindblad master equation supports a unique steady state configuration \cite{DNunique} and the generator $\mc L$ is weakly symmetric under the action of a finite group $G$, whose elements permute the lattice sites. Notice, however, that both the dimension and the structure of such group do depend on the number of spin $N$. For $N=2$ the system is symmetric under the action of $G=\mathbb{Z}_2$, and contains only two elements: the idenitity operator and a single reflection operator that swaps the lattice sites. For $N\geq 3$, due to the periodic boundary conditions, the lattice structure is equivalent to a polygon having $N$ sides. Therefore, in this latter case, the symmetry group is the dihedral group $D_{2N}$, which contains $2N$ elements \cite{hamermesh}: the $N$ rotations (corresponding to lattice traslations) belonging to the subgroup $\mc T$ of $G$, and $N$ reflections belonging to the subset $\mc R$ of $G$. In particular, notice that in the present case the symmetry group is given by the union of such two sets, that is 
\begin{equation}
G=\mc T \cup \mc R
\end{equation}
As a consequence, specifying a matrix representation for the elements of both $\mc T$ and $\mc R$ is equivalent to provide a matrix representation of the elements of $G$. To this purpose, in the case of $N$ spins, since the eigenstates of the total magnetization $\vert \bm{\sigma}\rangle$ can be mapped by means of an isomorphism $\mc F$ into a vector $\vec{v}_{\bm{\sigma}}$, that is
\begin{equation}\label{eq:vectors}
\mc F(\vert \bm{\sigma}\rangle)= \vec{v}_{\bm{\sigma}},
\end{equation}
with
\begin{equation}
\vec{v}_{\bm{\sigma}}\equiv\left(\sigma_1,\,\sigma_2,\,\cdots,\,\sigma_N\right)\quad \sigma_j=[-1,\,1],
\end{equation}
it is straightforward to see that such two subsets are completely determined by two operators, $T$ and $R$, whose action on a generic $\vec{v}_{\bm{\sigma}}$ is given by
\begin{equation}\label{eq:T_gen}
T\vec{v}_{\bm{\sigma}}=\left(\sigma_N,\,\sigma_1,\sigma_2,\cdots,\sigma_{N-2},\sigma_{N-1}\right),
\end{equation}
and by
\begin{equation}\label{eq:R_gen}
R\vec{v}_{\bm{\sigma}}= \left(\sigma_N,\,\sigma_{N-1},\sigma_{N-2},\cdots,\sigma_2,\,\sigma_{1}\right).
\end{equation} 
In particular, the $N$ rotations correspond to the elements of the following set
\begin{equation}\label{eq:rotations}
\mc T = \left\{T^{k},\,k \in [1,\,N]\right\},\quad (T^N=\bm{1})
\end{equation}
and the action of the $N$ reflections is instead given by the action of the elements of the following set
\begin{equation}\label{eq:reflections}
\mc R = \left\{ R\, T^{k},\,k \in [1,\,N]\right\},
\end{equation}
with $T$ and $R$ being the matrices defined by Eq \ref{eq:T_gen} and Eq. \ref{eq:R_gen} respectively. In particular, by means of Eqs. \ref{eq:T_gen} and \ref{eq:R_gen} one can easily compute the number orbits associated to the action of the symmetry group $G$ on the set $\left\{\vert \bm{\sigma}\rangle \langle \bm{\eta}\vert\right\}$, that is $\mbox{dim}[\mc I_{G}]$. A comparison between the number of variational parameters for $\alpha=1$ and $\alpha= 2$ used in the neural networks (we recall that we set $\beta=\alpha$), the dimension of $\mc I_G$ and  $D \times D$, for different values of $N$ is reported in Tab. \ref{tab:parameter_counting}.
\begin{table}[h]
\begin{tabular}{|c|c|c|c|c|}
\hline
$N$ & $N_p(\alpha=1)$ & $N_p(\alpha=2)$ & $\mbox{dim}[\mathcal{I}_G]$ & $D \times D$  \\\hline
$2$   &      14    &  26   &   10        &         16 \\
$3$   &      27    &  51   &   20        &         64  \\
$4$   &      44    &  84   &   55        &         256 \\
$5$   &      65    &  125  &  136        &         1024 \\
$6$   &		 90	   & 174   &	430		 &		   4096\\
$7$   &      119   &  231  &   1300      &         16384\\
$8$   &      132   & 296   &   4435      &         65536\\
\hline
\end{tabular}
\caption{A comparison between $N_p(\alpha)=(2\alpha+1)N+2\alpha\,N^2$ for $\alpha=1$ and $\alpha=2$, $\mbox{dim}[\mathcal{I}_G]$  and $D \times D = 2^N \times 2^N$ for a system of $N$ spins having the same symmetry group of the dissipative XYZ model in $d=1$.}\label{tab:parameter_counting}
\end{table} 

\subsection{Evolution of neural networks}\label{sec:evolution_method}
In this section we describe the optimization scheme exploited in our numerical simulations. Similarly to what has been discussed in \cite{open1,open2} and summarized in Sec \ref{sec:variational_opt_RBM}, we performed the numerical optimization of the neural networks by looking for the optimal sequence of variational vectors that best approximate the actual time evolution. Nevertheless, our scheme differs is some aspects from those described in \cite{open1,open2}. As mentioned in \cite{open1}, the matrix $S$ can be non-invertible. As a consequence, it might happen that the system in Eq. \ref{eq:linear_system} has no solution. In order to avoid this issue, we use a scheme based on the \emph{Moore-Penrose} pseudoinverse matrix $S^+$. Indeed \cite{pseudoinverse}, given any linear system involving a coefficient matrix $S$ and a target vector $\vec{f}$, we have that the pseudoinverse matrix $S^+$, which is \emph{always} defined, is such that 
\begin{equation}
\left\vert\left\vert S \frac{d\vec{\chi}}{dt}-\vec{f}\,\right\vert\right\vert_{2}\geq \vert\vert S\vec{z}-\vec{f}\vert\vert_{2},\quad\mbox{with}\quad\vec{z}=S^+\vec{f},
\end{equation}
with $\vert\vert \cdot \vert\vert_2$ denoting the standard Euclidean norm. In other words, even when $S$ is not invertible and the system has no solution, the pseudoinverse $S^+$ can be exploited to determine a least-squares solution of the linear system that can be used to update the variational vector. In addition, if $S$ is invertible with inverse given by $S^{-1}$, then $S^+=S^{-1}$. In particular, similarly to \cite{open2}, we used a matrix $S$ and a vector $\vec{f}$ having the following components
\begin{equation}\label{eq:my_S_matrix}
(S)_{i\,j}=\langle \rho_{\vec{\chi}}\vert O^{\dagger}_i\,O_j\vert \rho_{\vec{\chi}}\rangle+ \langle \rho_{\vec{\chi}}\vert O^{\dagger}_j\,O_i\vert \rho_{\vec{\chi}}\rangle,
\end{equation}
and 
\begin{equation}
(\vec{f})_i= \langle \rho_{\vec{\chi}}\vert O^{\dagger}_i M_{\mc L}\vert \rho_{\vec{\chi}}\rangle+ \langle \rho_{\vec{\chi}}\vert M^{\dagger}_{\mc L}\,O_i\vert \rho_{\vec{\chi}}\rangle
\end{equation}
Therefore, starting from a neural network representation with variational parameters having real and imaginary part randomly inizialized in the interval $[-0.01,\,0.01]$, at each iteration of the algorithm we first determined pseudo-inverse of the matrix defined in Eq.\ref{eq:my_S_matrix}, and then updated the variational vector as follows
\begin{equation}
\vec{\chi}_{(n+1)}=\vec{\chi}_{(n)} + \Delta t(\mathcal{N})S^+ \vec{f},
\end{equation}
with $\mathcal{N}= \vert\vert S^+ \vec{f}\,\vert\vert_{2}$ and with
\begin{equation}\label{eq:stop_criterion}
\Delta t(\mathcal{N})=
\left\{
\begin{split}
& \Delta t / \mathcal{N} \quad\mbox{if}\quad \mathcal{N} \geq 1\\
& \Delta t  \quad\mbox{if}\quad \mathcal{N} < 1\\
\end{split}\right.
\end{equation}
In particular, we do believe that the rule provided in Eq. \ref{eq:stop_criterion} overcomes the issue of determining the optimal value of $\nu$ mentioned in \cite{open1}, which in that context plays the same role as $\Delta t$ in ours.\\

\subsection{A performance comparison: RBM vs Invariant Ansatz}\label{sec:performance_comparison}
\begin{figure}
\includegraphics[scale=0.6]{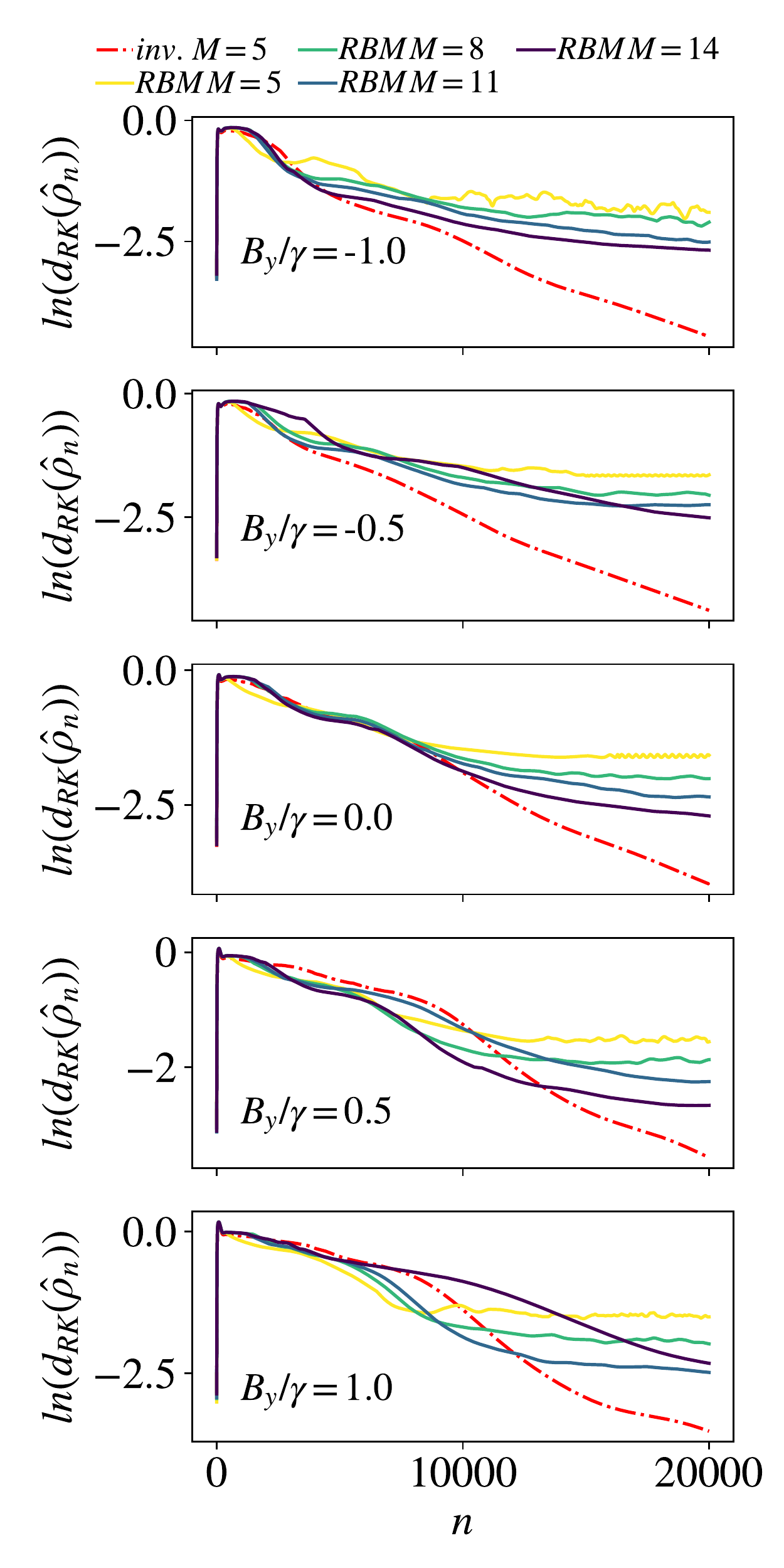}
\caption{Behavior of $ln(d_{RK}(\hat{\rho}_n))$ (see Eq. \ref{eq:distance_definition}) as a function of the number of iterations $n$ for $N=5$,  $J_x/\gamma = 1.4$, $J_y/\gamma = 2.0$, $J_z/\gamma = 1.0$, $B_x/\gamma = -1.0$, $B_z/\gamma=0.1$ and for the different values of $B_y/\gamma$ specified in the panels. The different curves in each panel correspond to a different neural network representation. The dotdashed red curve correspond to the invariant Ansatz defined in this paper. The trajectories labeled by $RBM$ correspond to neural networks having the form provided in Eq. \ref{eq:rbm_elements} and having the number of hidden spins $M$ specified in the legend on top of the figure.}\label{fig:distance_comparison}
\end{figure}

In this section we make a direct comparison between the expressive power of the RBM defined in Eq. \ref{eq:rbm_elements} and the invariant Ansatz defined in Eq. \ref{eq:symmetrical_ansatz}. We begin our analysis by considering the behavior of the components of the mean magnetization 
\begin{equation}
\langle\hat{M}_{k}\rangle_{n}=\frac{1}{N}\sum_{j=1}^{N}\mbox{Tr}\left[\sigma^{k}_{j}\,\hat{\rho}_n\right],
\end{equation}
\begin{figure}
\includegraphics[scale=0.7]{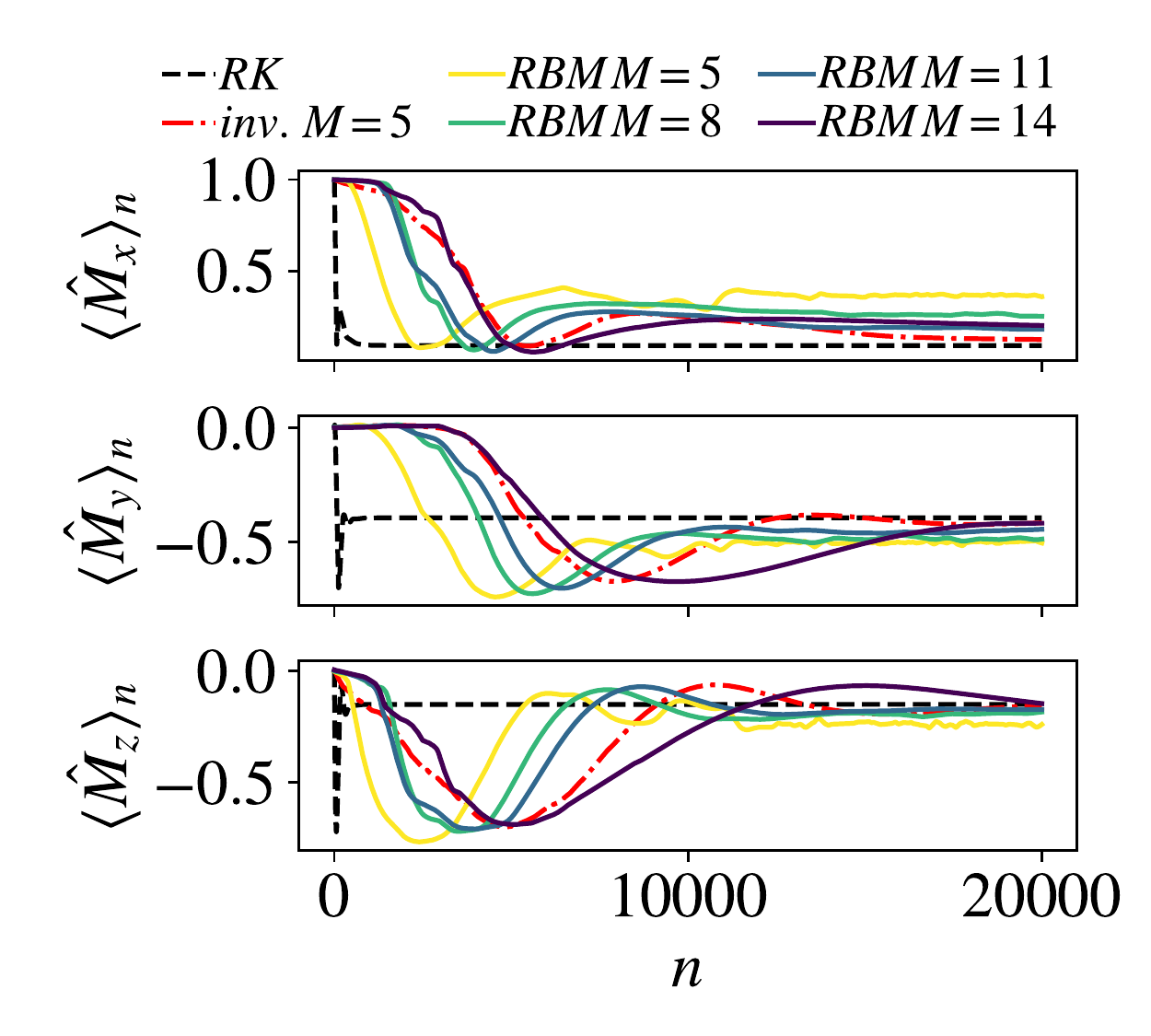}
\caption{Behavior of the components of the mean magnetization as a function of the number of iterations $n$ for $N=5$,  $J_x/\gamma = 1.4$, $J_y/\gamma = 2.0$, $J_z/\gamma = 1.0$, $B_x/\gamma = -1.0$, $B_y/\gamma=1.0$, $B_z/\gamma=0.1$ and $\gamma = 1$. The different curves correspond to: a fourth-order Runge-Kutta evolution with time step $\Delta t = 10^{-2}$ (black dashed line labeled by RK); the variational optimization of our Ansatz for $M=L=5$ (dash-dotted red curve labeled by $inv.\,M=5$); the variational optimization of four RBM having a number of hidden spin $M=5,\,8,\,11,\,14$ (labeled by $RBM$ followed by the corresponding number of hidden spins).}\label{fig:spin5_By1.0}
\end{figure}
over three different types of sequences of density operators $\{\hat{\rho}_n\}$: a sequence corresponding to a standard fourth-order Runge-Kutta evolution with time step $\Delta t=10^{-2}$; a sequence corresponding to the variational optimization of the invariant neural network Ansatz defined in this paper; several sequences corresponding to the trajectories associated to the variational optimization of RBMs having a different number of hidden spins. In particular, data corresponding to the neural networks have been obtained by using $M=L$ and by means of the strategy described in Sec \ref{sec:evolution_method}. Some results obtained for $N=5$ are shown in Fig. \ref{fig:spin5_By1.0}. There we show the behavior of the three components of the mean magnetization as a function of the number of iterations $n$ along the trajectories defined here above, and in correspondence of $J_x/\gamma = 1.4$, $J_y/\gamma = 2.0$, $J_z/\gamma = 1.0$, $B_x/\gamma = -1.0$, $B_y/\gamma=1.0$ and $B_z/\gamma=0.1$.\\
As it is possible to see, if one focuses on the trajectories labeled by $RBM$, coherently with the conclusions drawn in the previous papers, while increasing the number of hidden spins $M$ the RBM Ansatz becomes more and more expressive. Indeed, while increasing $M$ passing from $M=5$ to $M=14$ it is possible to notice that the RBM trajectories get closer and closer to the RK trajectory for large $n$, meaning that the optimization strategy is driving the neural network towards a state in the vicinity of the actual steady state. In addition, such results also suggest that our variational scheme based on the Moore-Penrose pseudoinverse and having a non-constant time step can be used for the variational optimization of a RBM.\\
Let us now pay attention to the two trajectories corresponding to $M=5$, that is $``inv.\, M=5"$ and $``RBM\,M=5"$. Such two curves describe respectively the variational optimization of our invariant Ansatz and that of a RBM in correspondence of the \emph{minimum} number of variational parameters that can be used in the neural network representation when assuming a $\hat{\rho}_{\vec{\chi}}$ such the one defined in Eq. \ref{eq:rbm_elements}. As it is possible to see, data suggest that the invariant Ansatz provides a better approximation of the steady state configuration. In addition, data shows that the symmetrization procedure can lead to the definition of a neural network representation which is also more expressive than a RBM Ansatz having an higher number of hidden spins $M$. This is already suggested by the data in Fig. \ref{fig:spin5_By1.0}. Nevertheless, for the sake of clarity, please pay attention to Fig. \ref{fig:distance_comparison}. There we show the behavior of the following quantity
\begin{equation}\label{eq:distance_definition}
ln(d_{RK}(\hat{\rho}_n))=ln(\vert \vert\, \hat{\rho}^{RK}_n-\hat{\rho}_n\, \vert \vert),
\end{equation}
which provides information about the distance between the RK trajectory and those associated to the variational optimization of the neural networks considered in this paper. In particular, notice that, similarly to what has been reported in Fig. \ref{fig:spin5_By1.0}, in all the cases considered in Fig. \ref{fig:distance_comparison} the RK algorithm can be safely considered at convergence for $n > 1000$, that is $\hat{\rho}^{RK}_n=\hat{\rho}_{ss}$ for $n > 1000$. As a consequence, data shown in Fig. \ref{fig:distance_comparison} provide information about the expressive power of different neural networks (please compare Eq. \ref{eq:distance_definition} to the left-hand side of Eq. \ref{eq:expressive_power}).\\
In each panel the curve labeled by $inv. M=5$ describes the distance between the RK trajectory ($\hat{\rho}^{RK}_n$) and the trajectory corresponding to the variational optimization of the invariant Ansatz for $M=5$ after the same number of steps $n$. The curves labeled by $RBM$ describe instead the behavior of the natural logarithm of the distance between the RK trajectory and those obtained by optimizing a RBM Ansatz for different values of the parameter $M$.
In particular, the sequences of density operators that lead to the mean magnetizations reported in Fig. \ref{fig:spin5_By1.0}, lead to the curves reported in the bottom panel of Fig. \ref{fig:distance_comparison}. As expected, for all the values of $B_y/\gamma$ considered in this section, while increasing the number of hidden spins the RBM Ansatz becomes more and more expressive, and the corresponging $ln(d_{RK}(\hat{\rho}_n))$ becomes more negative for large values of $n$. Nevertheless, at first glance one also notices that, in all the cases, the invariant Ansatz has a considerably higher expressive power, as suggested by the behavior of the dash-dotted curve in each panel. \\

\begin{figure}[htb]
\includegraphics[scale=0.55]{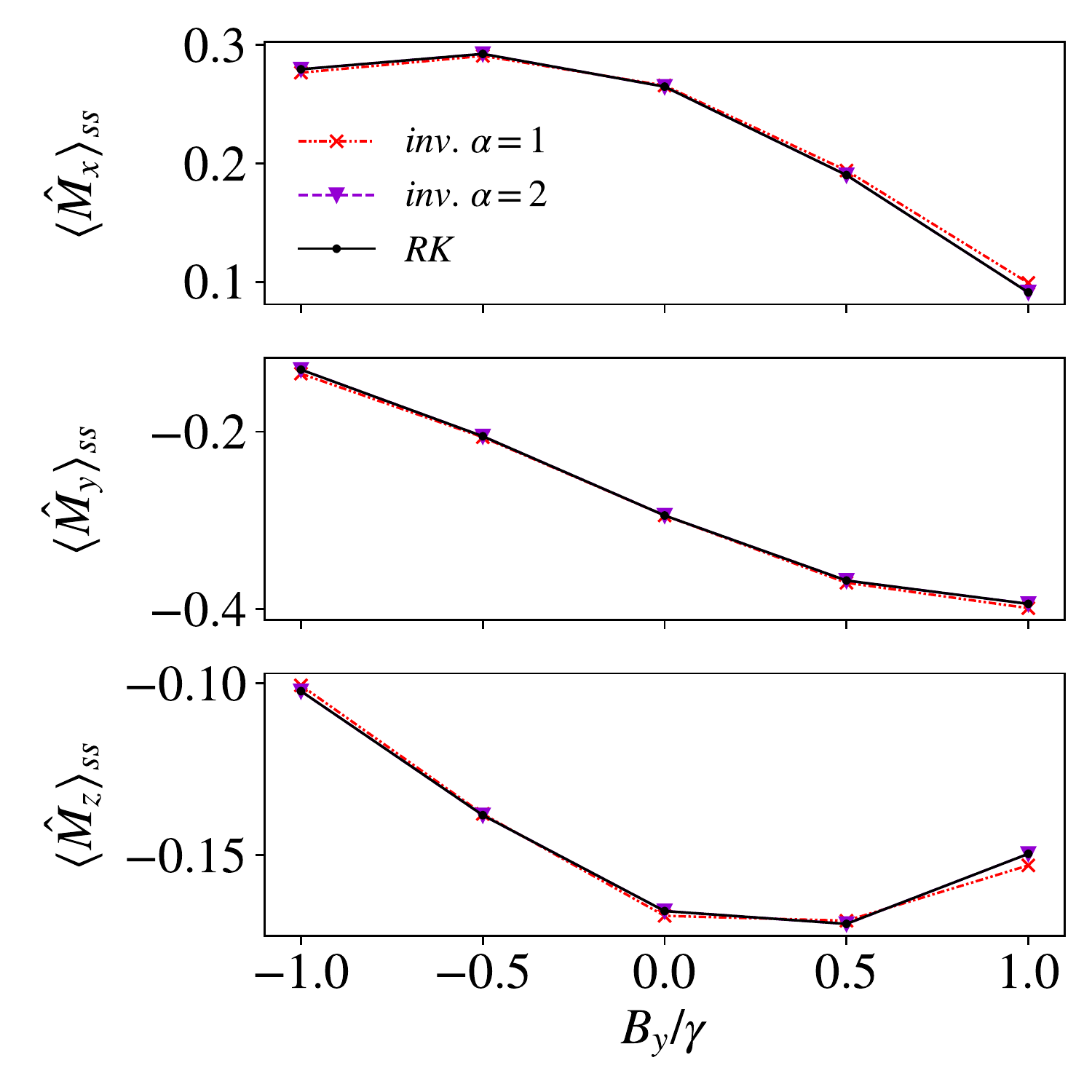}
\caption{Behavior of the steady state magnetizaion $\langle \hat{M}_{k}\rangle_{ss}$ ($k=x,\,y,\,z$) for the $N=5$ case, $J_x/\gamma = 1.4$, $J_y/\gamma = 2.0$,$J_z/\gamma = 1.0$, $B_x/\gamma = -1.0$, $B_z/\gamma=0.1$ and for different values of the magnetic field in the $y$ direction. The black solid curve corresponds to the steady state values obtained with the RK method. The red crosses and violet triangles correspond to the steady state values obtained optimizing the our neural network Ansatz, for different values of the parameter $\alpha$ (see the legend in the top panel). }\label{fig:steady_state_magnetization_spin5}
\end{figure}

Furthermore, similarly to what happens when using a RBM, if one increases the density of hidden spins $\alpha=M/N$ used in the representation, the invariant Ansatz becomes more expressive. To this purpose, please pay attention to Fig. \ref{fig:steady_state_magnetization_spin5} where we report the steady state mean values obtained by means of the RK method and by optimizing two invariant neural networks having $\alpha=1$ and $\alpha=2$, that is $M=5$ ($\alpha=1$) and $M=10$ ($\alpha=2$) respectively. As it is possible to see, there is a good agreement between the numerical results obtained by means of the different schemes. In particular, as expected, by increasing the number of variational parameters the numerical results obtained exploiting our approach become more and more accurate in capturing the steady state features (the violet curve corresponding to $\alpha=2$ is barely distinguishable from the black one corresponding to the RK results).
\subsection{Effective time evolution and convergence towards the steady state}\label{eq:effective_time_evolution}
In this last section we report some other numerical results describing the convergence towards steady states for $N\in [4,\,8]$. In particular, here focus on the results obtained by performing a variational optimization of an invariant Ansatz for $\alpha=1$ and $\alpha=2$. Indeed, similarly to what has been shown in the previous section, in most of the cases addressed in this section (especially for $N>5$), we observed that such number of variational parameters was not sufficient to obtain a good approximation of the steady state configurations when optimizing a standard RBM. Obviously, this does not prove that, for any possible choice of the number of spins $N$ and in correspondence of any possible point of the parameter space, the optimization of an invariant Ansatz having the same number of variational parameters as the ones considered in this section leads always to the determination of the steady state with an extremely high accuracy. Nevertheless, we do believe that the fact that in all the cases addressed in the present paper the results obtained for $\alpha=1$ provide a good approximation of the steady state mean values is a strong indication of the fact that, in general, it could be more convenient to optimize the invariant neural network Ansatz obtained by means of our symmetrization technique than to optimize a standard RBM.\\ 

\begin{figure}[h!]
\centering
\includegraphics[scale=0.7]{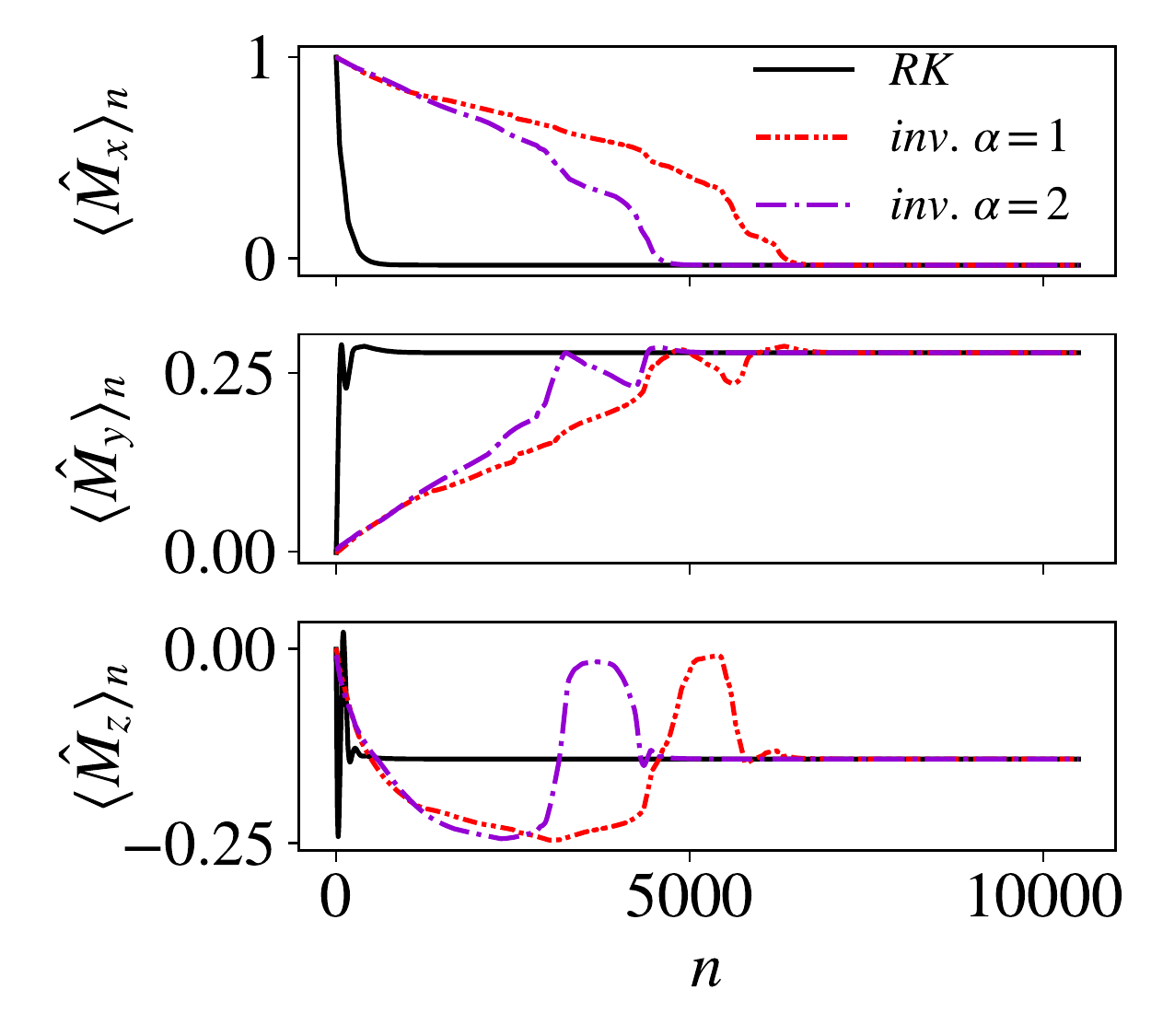}
\caption{Behavior of the components of the mean magnetization as a function of the number of iterations $n$ for $N=4$,  $J_x/\gamma = 1.3$, $J_y/\gamma = 0.1$, $J_z/\gamma = 1$, $B_x/\gamma = 0.7$, $B_y/\gamma=0.3$, $B_z/\gamma=0.1$. The different curves correspond to a fourth-order Runge-Kutta (RK) evolution with time step $\Delta t = 10^{-2}$, and to the variational optimization of our Ansatz for $\alpha=1$ ($inv.\,\alpha=1$) and for $\alpha=2$ ($inv.\,\alpha=2$). We used $\alpha=\beta$ in all the simulations.}\label{fig:spin4}
\end{figure}

Some results for $N=4$ are reported in Fig. \ref{fig:spin4}. There we show data for $J_x/\gamma = 1.3$, $J_y/\gamma = 0.1$, $J_z/\gamma = 1$, $B_x/\gamma = 0.7$, $B_y/\gamma=0.3$ and $B_z/\gamma=0.1$. As in the previous section, we use the trajectory generated by means of a RK evolution with fixed time step $\Delta t = 10^{-2}$ as a reference. 

\begin{figure}[h!]
\includegraphics[scale=0.6]{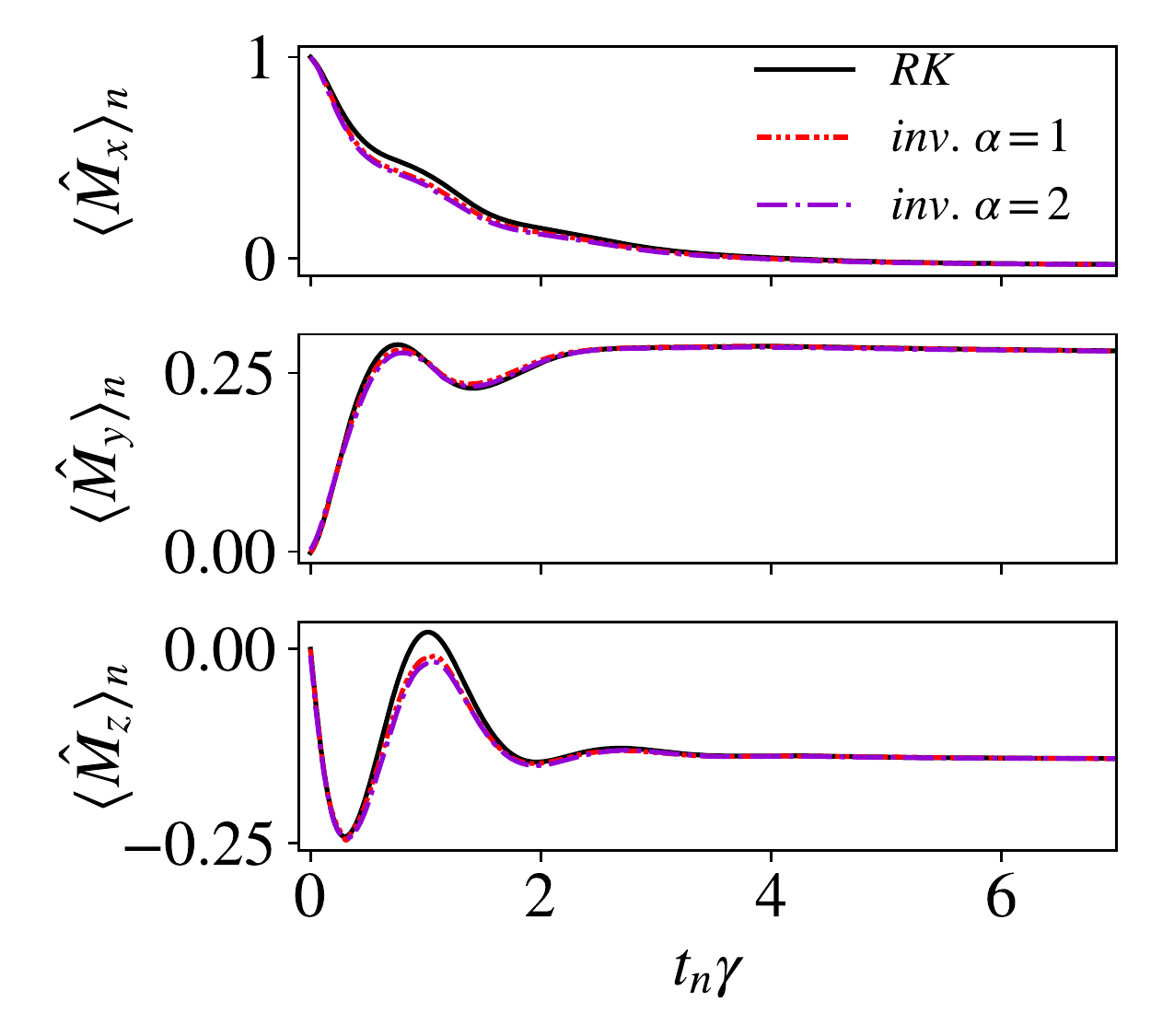}
\caption{Behavior of the components of the mean magnetization as a function of the physical time $t_n\gamma$ for $N=4$. The other parameters and the different curves are defined in Fig. \ref{fig:spin4}.}\label{fig:spin4_time}
\end{figure}
At first glance, one can notice that the trajectories corresponding to the variational optimization converge for sufficiently large $n$ to the RK trajectories. In particular, although the many trajectories display a similar global behavior (they have a similar shape), they do not overlap each other. Such fact is related to the prescription provided in Eq. \ref{eq:stop_criterion}. Indeed, as shown in Fig. \ref{fig:spin4_time}, if one uses the physical time instead of $n$, that is $t_{n+1}=t_n+\Delta t(\mathcal{N})$, it is possible to see that there is a good agreement between the trajectory in the parameter space determined by the variational optimization in the two cases considered and the RK evolution. However, as discussed in Ref. \cite{open2}, we do stress that having a perfect superposition with the actual time evolution is not necessary to determine the steady state.\\
\begin{figure}[h!]
\begin{center}
\subfigure[]{\includegraphics[scale=0.5]{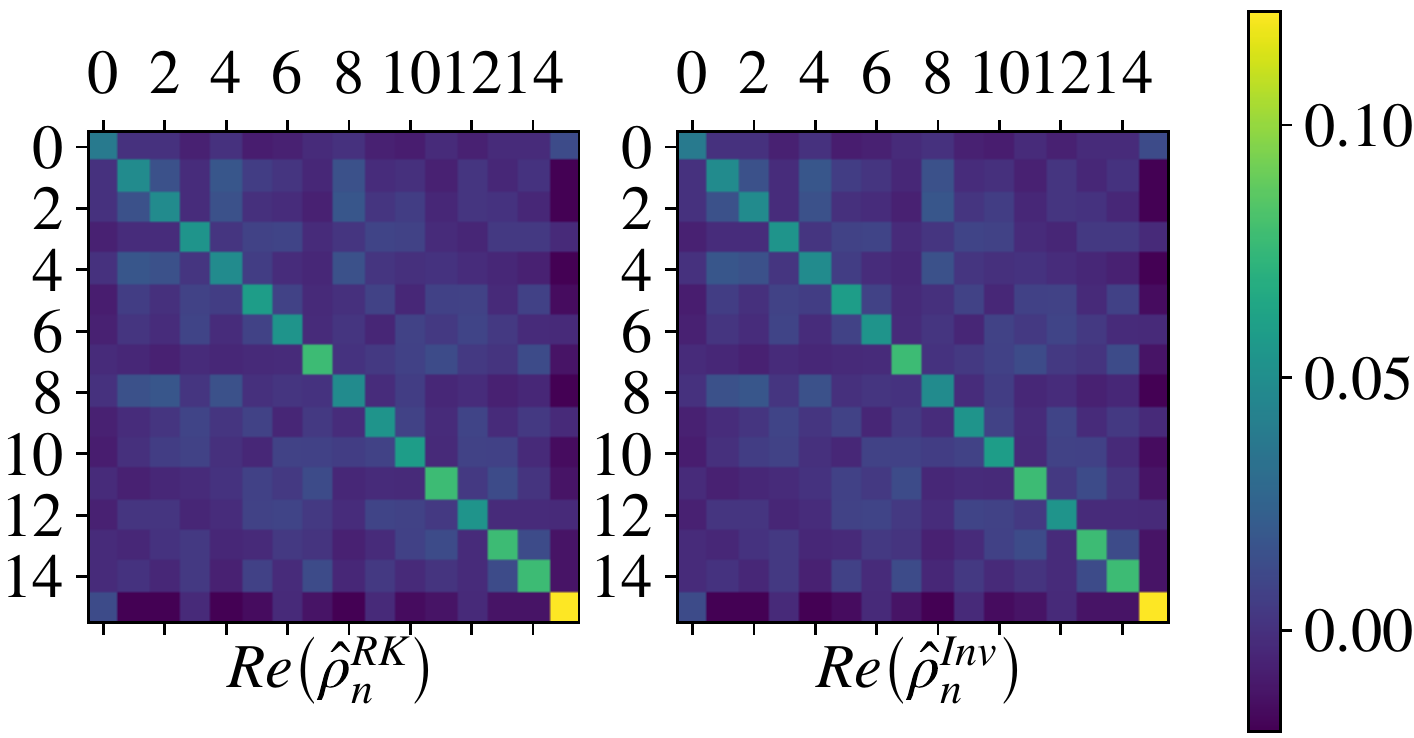}\label{fig:realPart_spin4}}
\subfigure[]{\includegraphics[scale=0.5]{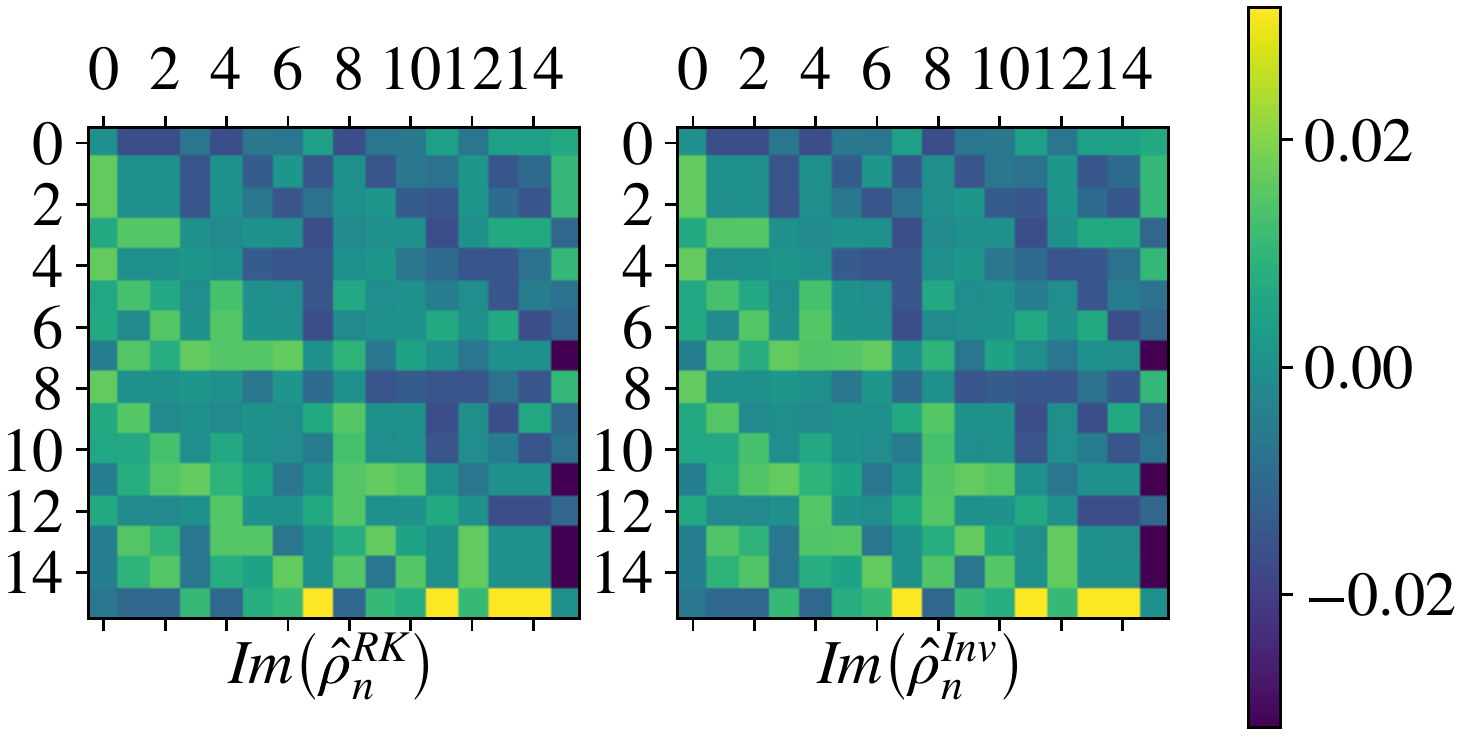}\label{fig:imagPart_spin4}}
\end{center}
\caption{System configurations obtained with the RK method ($\hat{\rho}^{RK}_{n}$) and by means of the variational optimization for $\alpha=1$ ($\hat{\rho}^{inv}_{n}$), for $N=4$ and $n=7000$. The other parameters are set as in Fig.\ref{fig:spin4}. Fig \ref{fig:realPart_spin4} describes the real part of the two matrices in correspondence of such point. Fig \ref{fig:imagPart_spin4} describes instead the imaginary part of such two matrices. In both cases, the label ``0" corresponds to the highest magnetization state. Therefore, the total magnetization of the basis states decreases while moving from the left to the right, and from the top to the bottom in each panel (the state labeled by ``15" denotes the lowest magnetization state with all the spins pointing down).}\label{fig:spin4_ss}
\end{figure}
The convergence towards the same limit is also suggested by the structure of the configurations obtained by means of the different schemes for large $n$. This is shown in Fig \ref{fig:spin4_ss}, where we compare the structures of the configurations reached at $n=7000$ and $N=4$, in the RK case and for a variational optimization with $\alpha=1$. As it is possible to see, both the real and imaginary part of such matrices, reported respectively in Fig. \ref{fig:realPart_spin4} and Fig \ref{fig:imagPart_spin4}, are extremely similar in the two cases.\\

\begin{figure}
\includegraphics[scale=0.6]{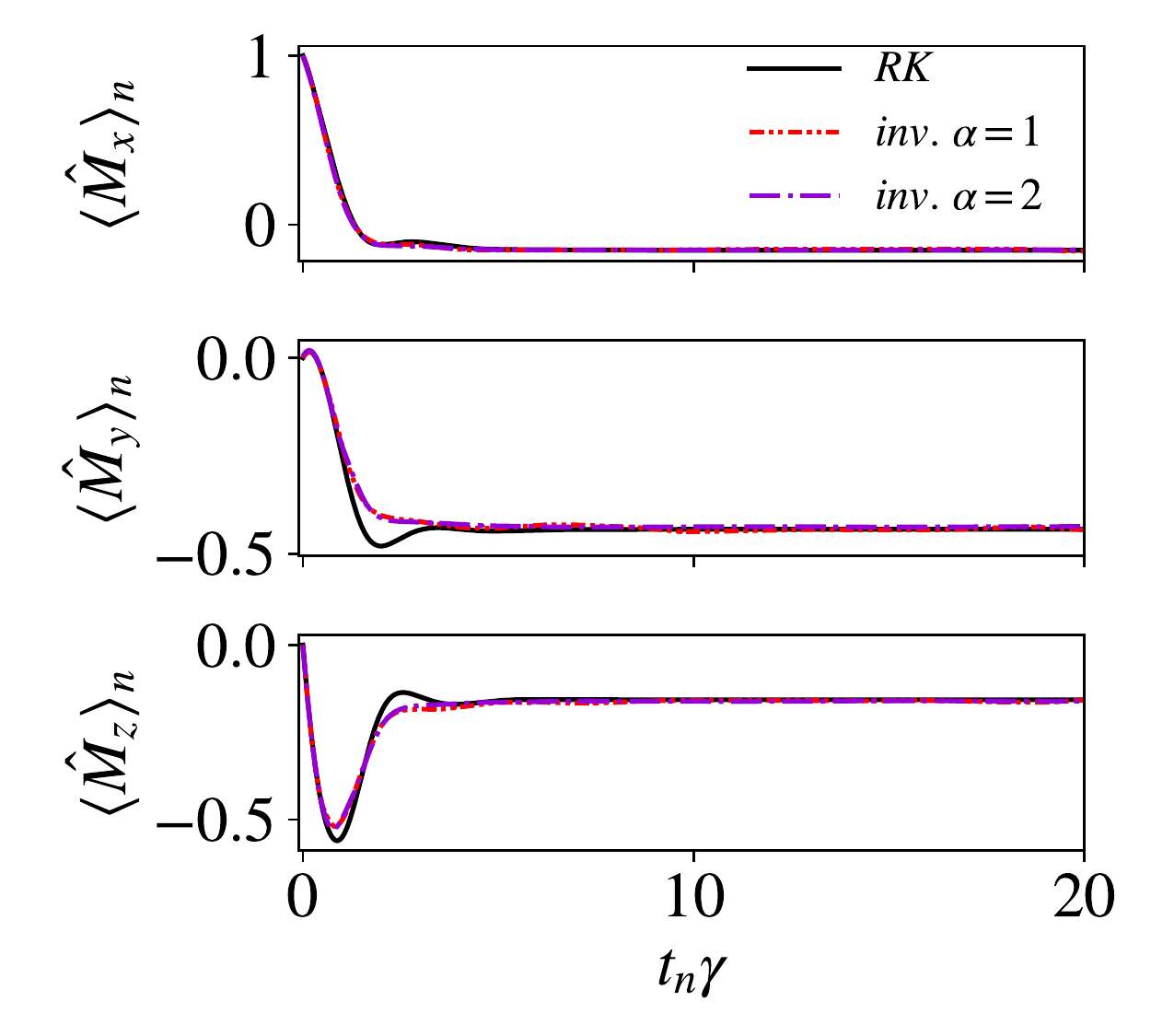}
\caption{Behavior of the components of the mean magnetization as a function of the physical time $t_n\gamma$ for $N=6$ ($\gamma=1.0$). The parameters used in the simulations are $J_x/\gamma=1.1$, $J_y/\gamma=0.5$, $J_z/\gamma=1.0$, $B_x/\gamma=-0.6$, $B_y/\gamma=0.2$ and $B_z/\gamma=0.1$ .}\label{fig:spin6}
\end{figure}

\begin{figure}
\includegraphics[scale=0.6]{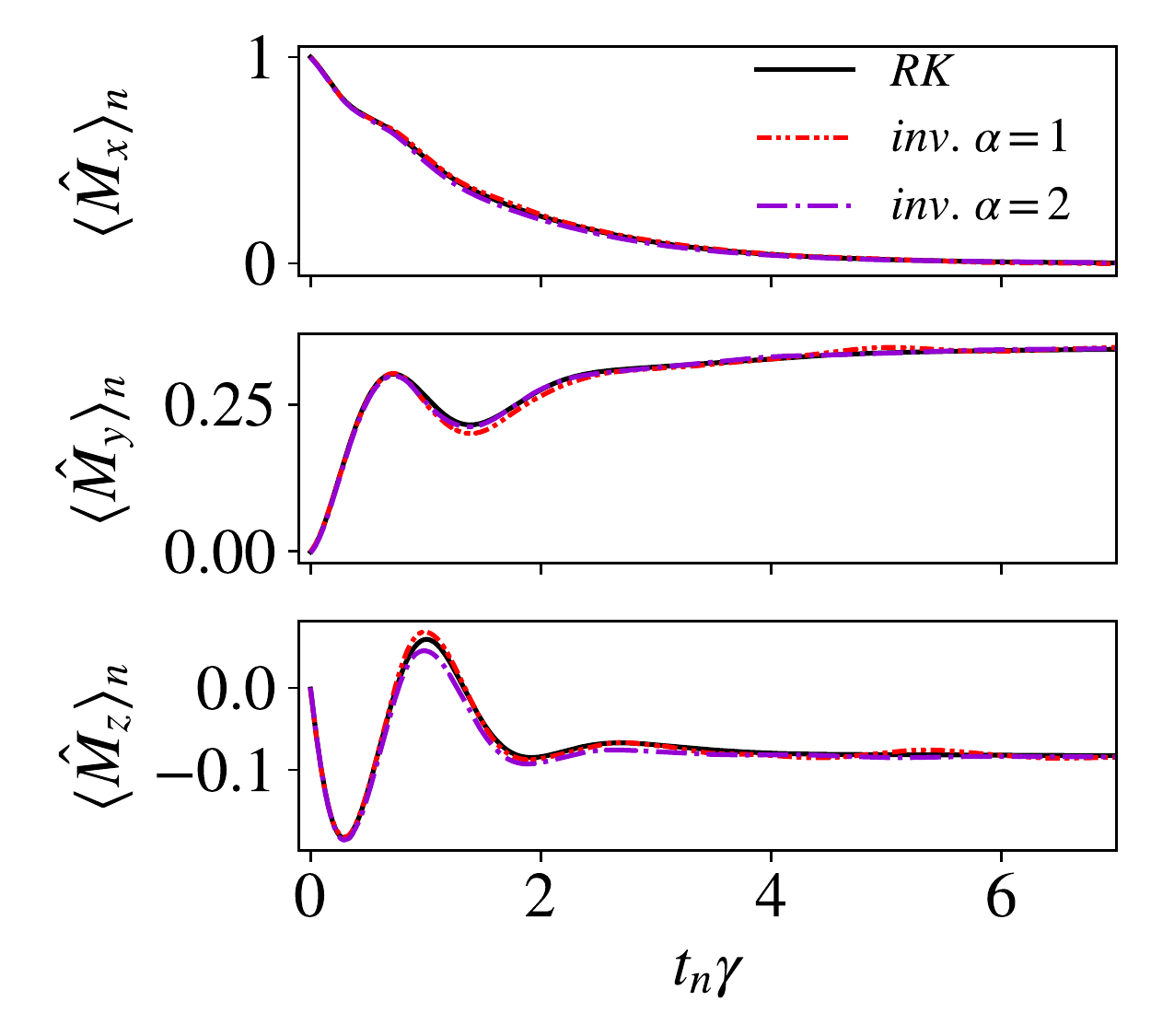}
\caption{Behavior of the components of the mean magnetization as a function of the physical time $t_n\gamma$ for $N=7$ ($\gamma=1.0$). The parameters used in the simulations are $J_x/\gamma=1.2$, $J_y/\gamma=0.3$, $J_z/\gamma=1.0$, $B_x/\gamma=1.2$, $B_y/\gamma=0.1$ and $B_z/\gamma=0.1$ .}\label{fig:spin7}
\end{figure}

\begin{figure}
\includegraphics[scale=0.6]{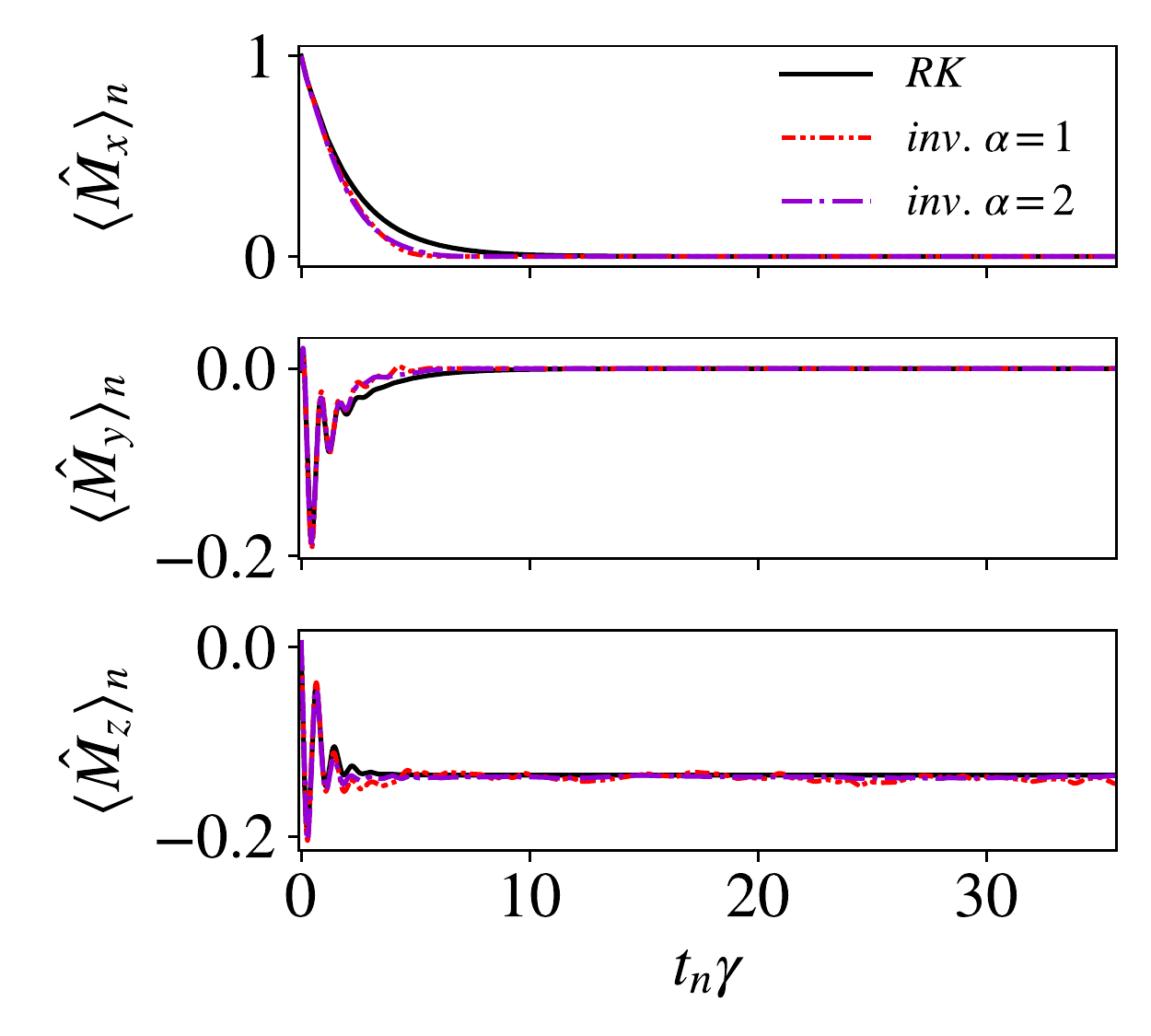}
\caption{Behavior of the components of the mean magnetization as a function of the physical time $t_n\gamma$ for $N=8$ ($\gamma=1.0$). The parameters used in the simulations are $J_x/\gamma=-1.1$, $J_y/\gamma=0.9$, $J_z/\gamma=1.0$, $B_x/\gamma=0$, $B_y/\gamma=0.3$ and $B_z/\gamma=0.1$ .}\label{fig:spin8}
\end{figure}

We conclude this section by showing some results obtained for $N=6,\,7,\,8$ in correspondence of other three different points in parameter space. In particular, we report their behavior directly as a function of the physical time. To this purpose, results for $N=6,\,7$ have been obtained by means of a time step $\Delta t=10^{-2}$. The results for $N=8$ have been obtained instead by setting the time step to $\Delta t= 5 \times 10^{-2}$. Results for such lattice chains are reported in Figs. \ref{fig:spin6}, \ref{fig:spin7} and \ref{fig:spin8} respectively.

As it is possible to see, although the trajectories in terms of the physical time do not overlap perfectly with those corresponding to the RK time evolution, we do see that the variational optimization of the invariant Ansatz defined in this paper is capable of capturing the magnetization features encoded into the steady state configurations. In particular, also in these three cases, as mentioned at the beginning of this section, it is possible to see that results for $\alpha=1$ provide a good approximation of the actual steady states in all the cases considered.\\

\section{Summary and Conclusions}\label{sec:summary_and_conclusions}
In this paper we considered an open quantum lattice system which is weakly symmetric under a finite group of permutations $G$ and whose time evolution is governed by a LGKS equation having a unique steady state configuration. In this case, it is known a priori that such configuration belongs to a subset $\mc I_G$ of the whole space of the density operators containing only states simultanously commuting with all the elements of the symmetry group $G$. Starting from a generic Restricted Boltzmann Machine (RBM), we exploited such information to define a $G$-invariant neural network Ansatz and provided an explicit theoretical proof of the stability of its symmetry properties under the action of any variational scheme similar to those discussed in Refs. \cite{open1,open2,open3,open4}. We discussed the validity of our approach by considering the one dimensional XYZ model in the presence of a uniform magnetic field and uniform dissipation. In particular, data reported in the present paper support the conjecture that the invariant neural network Ansatz is more expressive than a standard RBM. In other words, in correspondence of the same number of variational parameters $N_p$, we conjecture that the invariant neural network can approximate better than a standard RBM the states belonging to $\mc I_G$, thus the steady state. Nevertheless, in order to confirm such thesis, we think that further research is needed, especially for what concerns the relation between the number of parameters $N_p$ and the expressive power of neural networks. To this purpose, we mention \cite{glasser2018,chen2018,pastori2019} where the connection between neural networks and tensor networks has been explored in details. In addition, we do believe that some other issues deserve further investigation. For instance, it would be interesting to make a comparison between the performances obtained with  the different neural network representations appeared in literature so far, and to understand how the choice of the time step $\Delta t$ used to update the variational vector affects the convergence towards the steady state. We also believe that making a comparison between different evolution schemes would be of interest. That said, although here we investigated only a one dimensional open quantum systems, we stress that our approach can be easily exploited to characterize open quantum lattices in higher dimensions. Indeed, both the efficiency of variational schemes \cite{open1,open3,open4} and the power of weak symmetries \cite{DNsymm} have been exploited to determine the steady state configurations of two dimensional open quantum lattices. In addition, as discussed in Sec. \ref{sec:numerical_results}, our approach can be easily combined with stochastic methods. Therefore, we do believe that our approach will be used soon in combination with Monte Carlo techniques to characterise the asymptotic states of open quantum systems in $d \geq 2$.\\ 

We thank Dario Gerace and Giuseppe Clemente for helpful discussions. This work was supported by the Italian Ministry of University and Scientific Research (MIUR), PRIN Interacting Photons in Polariton Circuits INPhoPOL (2017P9FJBS$\_$001)

\end{document}